\documentclass[twocolumn,showpacs,amsmath,amssymb,prb]{revtex4}
\usepackage{graphicx,epsfig}

\newcommand{\bq}{\begin{equation}}
\newcommand{\eq}{\end{equation}}
\newcommand{\bqa}{\begin{eqnarray}}
\newcommand{\eqa}{\end{eqnarray}}
\newcommand{\bd}{\begin{displaymath}}
\newcommand{\ed}{\end{displaymath}}
\newcommand{\nn}{\nonumber \\}
\newcommand{\ij}{\langle i j \rangle}

\newcommand{\fbar}{f^*}

\newcommand{\dtau}{\partial_\tau}

\newcommand{\Ybar}{Y^*}

\newcommand{\ek}{\epsilon_k}

\newcommand{\gammabar}{\gamma^*}

\def\be     {\begin{equation}}
\def\ee     {\end{equation}}
\def\ba    {\begin{eqnarray}}
\def\ea    {\end{eqnarray}}
\def\bnn    {\begin{eqnarray*}}
\def\enn    {\end{eqnarray*}}

\begin{document}

\title{Slave rotor theory of antiferromagnetic Hubbard model}
\author{Ki-Seok Kim}
\email[Electronic address:$~$]{kimks@kias.re.kr}
\affiliation{School of Physics, Korea Institute for
Advanced Study, Seoul 130-012, Korea }
\author{Jung Hoon Han}\email[Electronic address:$~$]{hanjh@skku.edu}
\affiliation{Department of Physics and Institute for
Basic Science Research, \\
Sungkyunkwan University, Suwon 440-746, Korea} \affiliation{CSCMR,
Seoul National University, Seoul 151-747, Korea}
\date{\today}

\begin{abstract}
The slave-rotor mean-field theory of Florens and Georges is
generalized to the antiferromagnetic phase of the Hubbard model. An
effective action consisting of a spin rotor and a fermion is derived
and the corresponding saddle-point action is analyzed.
Zero-temperature phase diagram of the antiferromagnetic Hubbard
model is presented. While the magnetic phase persists for all values
of the Hubbard interaction $U$, the single-particle spectral
function exhibits a crossover into an incoherent phase when the
magnetic moment $m$ (and the corresponding $U$ values) lies within a
certain window $m_c < m < 1-m_c$, indicating a possible deviation
from the Hartree-Fock theory.
\end{abstract}

%\pacs{71.10.-w, 71.10.Hf, 71.27.+a, 11.10.Kk}

\maketitle

The Hubbard model has received a lot of attention theoretically as a
prototypical model for strong electron correlation in low
dimensions\cite{Hubbard}. In particular, much has been learned about
the quantum phases and the zero-temperature transitions between them
in this model. Brinkman-Rice (BR) theory\cite{BR} predicts a
metal-to-paramagnetic insulator transition at a finite Hubbard
interaction strength $U$, and further refinement by dynamical
mean-field theory (DMFT) in recent years supports the original BR
picture\cite{DMFT}.

Recently Florens and Georges (FG)\cite{FG} introduced an interesting
re-formulation of the Hubbard model. In their slave-rotor (SR)
theory, in similar spirit to the slave-boson representation, an
electron operator is decomposed as the product of a fermion and a
U(1) ``slave-rotor" operator. Analysis of the effective mean-field
action revealed that a quantum phase transition takes place between
metallic and paramagnetic insulating states as $U$ is increased
beyond a threshold value $U_c$. A good qualitative agreement between
FG's slave-rotor mean-field theory (SRMFT) with the DMFT predictions
was achieved while avoiding the use of heavy numerical machinery of
the latter method. More recently, SRMFT was employed in the
understanding of frustrated Hubbard model on the triangular
lattice\cite{leelee}.

Largely ignored in the above-mentioned
theories\cite{BR,DMFT,FG,leelee,DMFT-comment} is the spin
degrees of freedom responsible for the antiferromagnetism.
The nesting of the half-filled Fermi surface and the onset
of spin density wave are usually treated in the
Hartree-Fock (HF) theory while the strong effects of
Gutzwiller projection on the HF ground state are ignored. A
notable exception in the efforts to go beyond the
Hartree-Fock picture to understand the magnetic phase   is
given by the four-boson formulation of the Hubbard model by
Kotliar and Ruckenstein (KR)\cite{KR}. In KR's theory
strong on-site correlation effects as well as the magnetic
order were treated at the mean-field level. Later quantum
Monte Carlo study confirmed much of the mean-field
conclusions of KR\cite{hanke}. Both
references\cite{KR,hanke} focused on the overall phase
diagram, and the nature of the quasiparticle states in the
magnetic phase was not thoroughly discussed. A more recent
study on this subject\cite{BD} concluded that at strictly
zero temperature, as in the Hartree-Fock (HF) picture, the
quasiparticles in the antiferromagnetic phase of the
Hubbard model remain coherent.

Given the new machinery of SRMFT, we feel that it is worthwhile to
re-visit this issue in more detail. In this paper, we present a
natural extension of the original SRMFT theory that allows one to
treat the magnetic as well as the non-magnetic phase of the Hubbard
model. The saddle-point analysis of the effective action for the
half-filled model gives two phases, characterized by the
coherence/incoherence of the quasiparticles. Further consideration
of gauge fluctuation renders the phase transition into a crossover.
The magnetic ordering persists for all values of $U$ as in the HF
theory.
\\

We start with the Hubbard model
\be  H = - t\sum_{ij \sigma}c_{j\sigma}^{\dagger}c_{i\sigma}
-\mu\sum_{i\sigma} c^{\dagger}_{i\sigma}c_{i\sigma}   +
U\sum_{i}c_{i\uparrow}^{\dagger}c_{i\uparrow}
c_{i\downarrow}^{\dagger}c_{i\downarrow}  \ee
defined on the two-dimensional square lattice. The Hubbard-$U$ term
can be decomposed into charge and spin channels in standard fashion

\be
c_{i\uparrow}^{\dagger}c_{i\uparrow}c_{i\downarrow}^{\dagger}c_{i\downarrow}
= \frac{1}{4}[n_{i\uparrow} + n_{i\downarrow} ]^{2} - \frac{1}{4}[
n_{i\uparrow} - n_{i\downarrow} ]^{2}  \label{U1-decompose} \ee
with $n_{i\sigma} =c^{\dagger}_{i\sigma}c_{i\sigma}$. The latter
term in Eq. (\ref{U1-decompose}) was ignored in the study of
spinless states by FG.

Florens and Georges introduced a decomposition of the
electron operator as $ c_{i\sigma} = e^{-i\theta_i }
f_{i\sigma} $ where the rotor variable $e^{-i\theta_i}$
serves to keep track of the charge number at each site, as
the creation or annihilation of an electron is accompanied
by the phase change $e^{\pm i\theta_i}$. Here we propose
that \textit{a second rotor variable $e^{\pm i\phi_i}$ can
be introduced for the bookkeeping of spin numbers}. The
representation of the electron operator we propose is
\be c_{i\sigma} = e^{-i\theta_i - i\sigma \phi_i }
f_{i\sigma}. \label{KH-representation} \ee
The factor $\sigma = \pm 1$ is used to distinguish the
creation of up and down spins along the $\hat{z}$-axis.
Although we are focusing on the U(1) case here, a fully
SU(2)-invariant representation of the spin sector is also
possible.

After the substitution made in Eq. (\ref{KH-representation}), the
local charge $\sum_\sigma c^{\dagger}_{i\sigma} c_{i\sigma}$ and the
local $z$-spin $\sum_\sigma \sigma c^{\dagger}_{i\sigma}c_{i\sigma}$
appearing in the decomposition of the Hubbard term are replaced by
the conjugate operators of $(\theta_i, \phi_i)$, which we denote $(
L_i , M_i ) = (-i\partial/\partial\theta_i,
-i\partial/\partial\phi_i)$. The Hubbard Hamiltonian takes on the
expression
\ba  &&  H  = {U\over 4} \sum_{i} L_i^2 - {U\over 4} \sum_i M_i^2
-\mu\sum_{i\sigma} f^{\dagger}_{i\sigma}f_{i\sigma} \nn
&& - t\sum_{ij \sigma}f^{\dagger}_{j\sigma} e^{i(\theta_j - \theta_i
) + i\sigma (\phi_j - \phi_i ) } f_{i\sigma}  \ea
supplemented by the constraints $L_i = \sum_\sigma
f^{\dagger}_{i\sigma} f_{i\sigma}$ and $M_i = \sum_\sigma
\sigma f^{\dagger}_{i\sigma}f_{i\sigma}$ at every
site\cite{FG}. The corresponding action can also be derived
straightforwardly:
\bqa && L = \sum_{i\sigma}f_{i\sigma}^{*}(\partial_{\tau} -
\mu)f_{i\sigma} - t\sum_{ij\sigma} f_{j \sigma}^{*}e^{i(\theta_{j} -
\theta_{i})+i\sigma (\phi_j -\phi_i ) }f_{i\sigma} \nn && +
\sum_{i}\Bigl[\frac{U}{4}L_{i}^{2} - iL_{i}\partial_{\tau}\theta_{i}
+ i\lambda_{i}(L_{i} - \sum_\sigma {f}_{i\sigma}^{*}{f}_{i\sigma})
\Bigr] \nn && + \sum_{i}\Bigl[-\frac{U}{4}M_{i}^{2} -
iM_{i}\partial_{\tau}\phi_{i} + i\eta_{i}(M_{i} - \sum_\sigma \sigma
{f}_{i\sigma}^{*}{f}_{i\sigma}) \Bigr] . \nn \label{LM-action}\eqa
A pair of Lagrange multipler fields $(\lambda_i , \eta_i )$
has been introduced to impose the constraints. The original
theory of FG is based on this Lagrangian, without the terms
pertaining to the spin decomposition such as $M_i$,
$\phi_i$, and $\eta_i$. Our formulation thus generalizes
the scheme of FG in a natural way to include magnetic
order.

To proceed further, we ignore the terms pertaining to the charge
sector as they are already discussed by FG. Leaving out the terms
containing $(\theta_i, L_i, \lambda_i)$, the next step is to
integrate out $M_i$ from the action. Because of the negative norm in
$M_i^2$ (third line of Eq. (\ref{LM-action})), we use the following
identity of the Gaussian integration

\ba && \int_{-\infty}^\infty dM \int_{-\infty}^\infty d\eta ~~
e^{{1\over 2} M^2 + i a M - i\eta (M - s) } \nn && =
\int_{-\infty}^\infty d\eta  \int_{-\infty}^\infty dM ~~ e^{-
{1\over 2} M^2 - a M  - i\eta ( M + i  s )  }
\ea
to first re-write the action (\ref{LM-action}) with the
\textit{positive norm} for $M_i^2$. The integration over
$M_i$ of the modified action can then proceed to give

\ba && L = \sum_{i\sigma}f_{i\sigma}^{*}(\partial_{\tau} - \mu -
\sigma \eta_i )f_{i\sigma} + {1\over U}\sum_{i} (
 \eta_i - i\dtau\phi_i )^2  \nn
 && ~~~ - t\sum_{ij\sigma}
f_{j \sigma}^{*}e^{i\sigma (\phi_j -\phi_i )} f_{i\sigma} .
\label{spin-only-action}  \ea
The final manipulation involves the shift $\eta_i \rightarrow \eta_i
+ i\dtau \phi_i$ that gives

\ba && L = \sum_{i\sigma}f_{i\sigma}^{*}(\partial_{\tau} \!-\! \mu
\!-\! \sigma \eta_i \!-\! i \sigma \dtau \phi_i )f_{i\sigma} +
{1\over U}\sum_{i} \eta_i^2  \nn
&& ~~~ - t\sum_{ij\sigma} f_{j \sigma}^{*}e^{i\sigma (\phi_j -\phi_i
) }f_{i\sigma} . \label{spin-only-action2}  \ea
This concludes the formal derivation of the effective action for the
spin-ful Hubbard model within the slave-rotor framework. Without the
phase fluctuations in Eq. (\ref{spin-only-action2}) this effective
action at the saddle-point level is equivalent to the HF theory of
the Hubbard model.

To explore the consequences of phase fluctuations in the action
(\ref{spin-only-action2}), we replace $\eta_i$ by its saddle-point
value $\overline{\eta}_i$ given by

\be \overline{\eta}_i = {U\over2} \langle \sum_\sigma \sigma
\fbar_{i\sigma} f_{i\sigma} \rangle  \equiv {U\over2} m_i, \ee
which is identical to the saddle-point equation in the absence of
the phase fluctuation. We will write $\eta_i$ to refer to this
average value in the rest of the paper.

One can decompose the hopping term in Eq. (\ref{spin-only-action2})
by a pair of Hubbard-Stratonovich (HS) fields $(\alpha_{ij},
\beta_{ij})$\cite{leelee}, resulting in the effective Lagrangian

\bqa && L = L_{0} + L_{f} + L_{\phi} , \nn
&& L_{0} =
t\sum_{\ij}(\alpha_{ij}\beta_{ij}^{*}+\alpha_{ij}^{*}\beta_{ij} ) +
{1\over U}\sum_i \eta_i^2 , \nn
&& L_{f} = \sum_{i\sigma}f_{i\sigma}^{*}(\partial_{\tau} \!-\! \mu
\!-\! \sigma \eta_i \!-\! i \sigma \dtau \phi_i )f_{i\sigma} \nn &&
~~~~ - t\sum_{\ij} (\fbar_{j\uparrow}\beta_{ij}f_{i\uparrow} +
\fbar_{j\downarrow}\beta_{ij}^{*}f_{i\downarrow} + c.c. ) , \nn
&& L_{\phi} = - t\sum_{\ij} (e^{-i\phi_{j}}\alpha_{ij}e^{i\phi_{i}}
+ c.c. )  \label{KH-effective-action} .\eqa
At the saddle-point level, the HS parameters take on the average
values $ \alpha_{ij} =\langle \fbar_{i\uparrow}f_{j\uparrow} +
\fbar_{j\downarrow}f_{i\downarrow} \rangle$, $\beta_{ij} =\langle
e^{-i \phi_i } e^{i\phi_j } \rangle$.
Assuming real and uniform mean-field solutions $\alpha_{ij} =
\alpha$, $\beta_{ij} = \beta$, and staggered effective magnetic
fields $\eta_i = (-1)^i \eta$, $m_i = (-1)^i m$, we arrive at the
mean-field effective action:
\bqa L_{MF} &=& L_{f} + L_{\phi} , \nn
\nn
L_{f} &=& \sum_{i\sigma }f_{i\sigma}^{*}(\partial_{\tau} \!-\! \mu
\! -\! \sigma (-1)^i \eta )f_{i\sigma} - t\beta\sum_{ij\sigma}
\fbar_{j\sigma} f_{i\sigma} \nn
L_{\phi} &=& -2t\alpha \sum_{\ij} \cos [\phi_j - \phi_i ] - i \sum_i
m_i \dtau \phi_i  \nn
\label{mean-field-for-S} \eqa
We assumed the case of no broken time-reversal symmetry, $\langle
\dtau \phi_i \rangle = 0$. The fermion action is in the standard HF
form except for the renormalization of the bandwidth $t\rightarrow
\beta t$. At zero temperature the fermion sector thus always remains
in the magnetic phase with a gap to quasiparticle excitations. The
boson action, on the other hand, is the standard XY action modified
by the Berry phase term $i\sum_i m_i \dtau \phi_i$. We analyze each
of the mean-field actions derived in Eq. (\ref{mean-field-for-S}),
beginning with the fermion sector.

In analyzing the fermionic mean-field action we confine our
attention to half-filling for which $\mu=0$. The mean-field
conditions for $\eta$ and $\alpha$ at $T=0$ read
\be \eta = U \eta \sum'_k {1 \over E^f_k }, ~~ \alpha = {2\over zt}
\sum'_{k} \ek \times {\beta \ek \over E^f_k} .
\label{fermion-SC-eqs}\ee
Here $\ek = -t \sum_{j\in i} e^{i k \cdot (r_j - r_i ) }$ ($j$ runs
over all nearest neighbors of $i$) is the bare band in the absence
of exchange splitting introduced by non-zero $\eta$, and $z$ is the
lattice coordination number. The $k$-sum in both equations is over
the reduced Brillouin zone and $E^f_k = \sqrt{(\beta\ek )^2 +
\eta^2}$ is the fermionic energy with a gap set by $\eta$.

The bosonic action derived in Eq. (\ref{mean-field-for-S}) is
invariant under $m\rightarrow -m$ or $m\rightarrow 1-m$ followed by
a shift of one lattice spacing. The Berry phase term vanishes for
$m=0$ and $m=1$ leaving only the classical XY model with an ordered
phase $\phi_i = \phi_0 $ at zero temperature. One might thus expect
that small deviations such as $m\approx 0$ and $m \approx 1$ still
gives the ordered phase while $m \approx 0.5$ introduces enough
perturbation to induce phase disordering. We present a calculation
which confirms this expectation.

Following the treatment of FG\cite{FG} we introduce the uni-modular
field $Y_i$ to make the replacement $e^{i\phi_i } \equiv Y_i$. An
additional Lagrange multiplier $q_i$ is required to impose the
uni-modular constraint.
This extension allows us to examine the bosonic action at the
saddle-point level. The effective boson Lagrangian written in terms
of $Y_i$ reads

\ba
&& L_{Y} =  - t\alpha\sum_{ij} \Ybar_j Y_i + i\sum_i q_{i} (\Ybar_i
Y_i -1)  \nn && ~~~~ -{m\over2}\sum_i (-1)^i [\Ybar_i \dtau Y_i -
(\dtau \Ybar_i) Y_i ] \nn && = \sum'_{k\nu} \Ybar_{k\nu} \left(  q +
\alpha \epsilon_k \right) Y_{k\nu} + \sum'_{k\nu} \Ybar_{k+Q \nu}
\left( q -\alpha \epsilon_{k} \right) Y_{k+Q \nu} \nn && ~~~~ -m
\sum'_{k\nu} i\nu ( \Ybar_{k+Q \nu} Y_{k \nu} + \Ybar_{k \nu} Y_{k+Q
\nu} ) , \ea
with $Q=(\pi,\pi)$. In writing down the Fourier form we assumed a
uniform mean-field solution $i \langle q_i \rangle = q$. The bosonic
$k$-sum is also over the reduced Brillouin zone. The boson part can
be diagonalized using a pair of operators $(\gamma_{1k\nu},
\gamma_{2k\nu})$ related to $(Y_{k\nu} , Y_{k+Q \nu})$ by
\ba Y_{k\nu} &=& {1\over\sqrt{2}} (\cosh \theta_k - \sinh \theta_k )
(\gamma_{1k\nu} +\gamma_{2k\nu} ) \nn
Y_{k+Q \nu} &=&{1\over\sqrt{2}} (\cosh \theta_k + \sinh \theta_k )
(\gamma_{1k\nu} -\gamma_{2k\nu} ). \ea
After taking $\cosh 2\theta_k = q/E^b_k$, $\sinh 2\theta_k = \alpha
\ek /E^b_k$, and $E^b_k = \sqrt{q^2 - (\alpha \ek)^2}$, one gets

\ba && L_Y = -im\sum'_{k\nu} \nu (\gammabar_{1k\nu} \gamma_{1k\nu} -
\gammabar_{2k\nu} \gamma_{2k\nu} ) \nn &&  ~~~~+\sum'_{k\nu} E^b_k
(\gammabar_{1k\nu} \gamma_{1k\nu} +\gammabar_{2k\nu} \gamma_{2k\nu}
) .\ea
The boson spectrum $E^b_k$ is gapped if $q - \alpha{D}
> 0$ while $q - \alpha{D} = 0$ leads to the condensation of
$Y$. Here $D=zt$ is half the bare bandwidth. Two additional
relations are obtained at $T=0$ from the constraints $\langle
\Ybar_i Y_i \rangle =1$ and $\beta = \langle \Ybar_i Y_j \rangle$:

\be {m \over q} = \sum'_k {1 \over E^b_k  }, ~~~\beta =  {\alpha
\over zt}  \sum'_k { \ek^2 \over m} {1 \over E^b_k } .
\label{boson-SC-eqs}\ee
Solving Eqs.  (\ref{fermion-SC-eqs}) and (\ref{boson-SC-eqs})
simultaneously renders the self-consistent parameters
$(\alpha,\beta,m,q)$ for a given $U/D$. The Bose condensation occurs
when $q=\alpha D$, which gives $m_c \approx 0.69$.

%%%%%%%%%%%%%%%%%%%%%%%%%%%%%%%%%%%%%%%%%%%%%%%%%%%%%%%%%%%%%%%%%%%%%%%%%%%%%%%%%%%%%%%%%
\begin{figure}[h]
\begin{center}
\includegraphics[width=8cm]{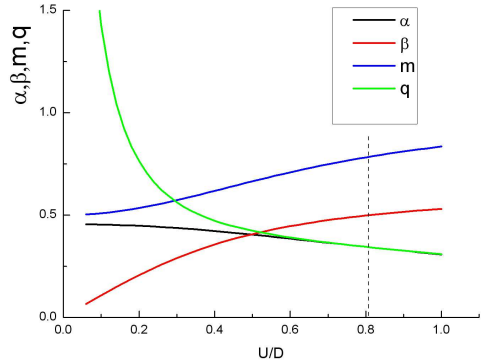}
\end{center}
\caption{(color online) Plot of mean-field parameters
$(\alpha,\beta,m,q)$ for $U/D$ obtained from solving Eq.
(\ref{MFE}). $q$ equals $\alpha D$ at $(U/D)_c \approx 0.81$
indicated by a dashed vertical bar. } \label{MF-vs-U}
\end{figure}
%%%%%%%%%%%%%%%%%%%%%%%%%%%%%%%%%%%%%%%%%%%%%%%%%%%%%%%%%%%%%%%%%%%%%%%%%%%%%%%%%%%%%%%%%

To get a better idea on the analytical structure of the set of
self-consistent equations obtained above, we first re-write Eqs.
(\ref{fermion-SC-eqs}) and (\ref{boson-SC-eqs}) as the integration
over the energy with a certain density of states $D(\epsilon)$, and
approximate it with a constant value, $D(\epsilon) = 1/(2D)$. The
mean-field equations are then given by

\ba  m={2\beta D/U \over \sinh[ 2\beta D/U]}, && \alpha =
{\sinh[4\beta D/U ] - 4\beta D/U \over 4\sinh^2 [2 \beta D/U ]} \nn
 q ={\alpha D  \over \sin[ 2 m \alpha D / q]}, && \beta = {4 m
\alpha D/q - \sin [4 m \alpha D/q ]  \over 8m \sin^2 [2 m\alpha D
/q]} .\label{MFE} \nn\ea
The Bose condensation occurs at $m_c = \pi/4 \approx 0.8$ ($(U/D)_c
\approx 0.81$), similar to the earlier estimate. The plot of all
four self-consistent parameters with varying $U/D$ using Eq.
(\ref{MFE}) is shown in Fig. \ref{MF-vs-U}.

The set of  mean-field equations (\ref{MFE}) yields
$m=1/2$, not $m=0$ as in the HF theory, when $U/D
\rightarrow 0$. We believe this is an artifact of the
replacement $e^{i\phi_i} \rightarrow Y_i$. The boson action
in Eq. (\ref{mean-field-for-S}), being equivalent to the XY
model, must show ordered phase at $m=0$ and $m=1$, and the
only chance of a disordered phase occurs for intermediate
$m$ values. Combining this general argument with the
mean-field analysis in terms of $Y$, we conclude that the
incoherent phase exists within the window $1-m_c \lesssim m
\lesssim m_c$ with $m_c \approx 0.8$. In this intermediate
phase the boson gap $\Delta_b = \sqrt{q^2 -(\alpha D)^2 }$
remains finite. Due to the gap, the spectral function of
the composite electron operator is
incoherent\cite{FG-comment}. The region outside this value
gives Bose condensation, or the coherent quasiparticles.
Meanwhile the magnetic sector remains ordered for all $U/D$
as in the HF theory. For comparison we recall that in the
paramagnetic Hubbard model\cite{FG}, SRMFT gave Bose
condensation for small-$U$: $ 0 < U/D < (U/D)_c$.

So far we performed  saddle-point analysis and obtained a mean-field
picture showing a second order phase transition (with the boson gap
as the order parameter) for the spin phase field $\phi_{i}$ in the
intermediate values of $U/D$. It is natural to ask the stability of
the mean-field picture against the gauge field $a_{ij}$ that appears
in the phase fluctuations of the hopping order parameters,
$\alpha_{ij} = \alpha{e}^{ia_{ij}}$ and $\beta_{ij} =
\beta{e}^{ia_{ij}}$, where $\alpha$ and $\beta$ are the mean field
values obtained before. It should be noted that the U(1) spin-gauge
field $a_{ij}$ is compact, thus allowing instanton
excitations\cite{Polyakov}.  From the seminal work of Fradkin and
Shenker\cite{FS_Instanton} we know that there can be no phase
transition between the Higgs and confinement phases due to instanton
proliferation, and only a crossover behavior is expected. In the
present problem the phase-coherent state corresponds to the Higgs
phase while the phase-incoherent state coincides with the
confinement phase. Applying Fradkin and Shenker's result to the
present problem, we conclude that the second order phase transition
turns into a crossover between the coherent and incoherent phases.
The magnetic order parameter, being a gauge-invariant quantity,
remains unaffected by the gauge fluctuation.

In summary we have developed an extension of the
slave-rotor theory of Florens and Georges to the
magnetically ordered phase by introducing a second rotor
variable pertaining to the spin degrees of freedom. On
performing a saddle point analysis we uncover an
incoherent-to-coherent crossover within the half-filled
antiferromagnetic phase of the Hubbard model at zero
temperature. The incoherent phase exists in the
intermediate values of $U/D$ between weak and strong
coupling limits. Given the common conception that the
magnetically ordered phase of the Hubbard model at $T=0$ is
well understood within the HF theory, the possibility we
suggest in this paper is tantalizing. The formalism
developed in this work may also be of use for understanding
other exotic magnetic systems.

HJH was supported by Korea Research Foundation through Grant No.
KRF-2005-070-C00044.


\begin{thebibliography}{9}
\bibitem{FG-comment} In FG's work as in ours, the Bose condensation marks
the coherence of quasiparticles. A finite boson gap implies that the
composite object's spectral function is incoherent. For an explicit
calculation of the electron spectral function see Ref.
\onlinecite{FG}.
\end{thebibliography}
\end{document}